\documentclass[10pt,twocolumn]{article}
\pdfoutput=1
\usepackage{graphicx}
\usepackage{SciAdvArXiv}

\usepackage{pdfpages}
\usepackage{times}
\usepackage{amsmath}
\usepackage{dsfont}
\usepackage{graphicx} 
\usepackage{capt-of}
\usepackage{color}
\usepackage{amsfonts}
\usepackage{amssymb}
\usepackage{multicol,caption,float}



\newenvironment{sciabstract}{%
\begin{quote} \bf}
{\end{quote}}

\title{\textsf{\textbf{A quantum Fredkin gate}}}

\author
{Raj B. Patel,$^{1,\ast}$ Joseph Ho,$^{1}$ Franck Ferreyrol,$^{1,2}$ Timothy C. Ralph,$^{3}$ \\
\& Geoff J. Pryde,$^{1,\ast}$\\
\\
\normalsize{$^{1}$CQC2T and Centre for Quantum Dynamics, Griffith University,}\\
\normalsize{Brisbane 4111, Australia}\\
\normalsize{$^{2}$ Laboratoire Photonique, Numerique et Nanosciences, Institut d'Optique,}\\
\normalsize{CNRS and Universit\'{e} de Bordeaux, Talence, France}\\
\normalsize{$^{3}$ CQC2T and School of Mathematics and Physics, University of Queensland, }\\
\normalsize{Brisbane 4072, Australia}\\
\normalsize{$^\ast$To whom correspondence should be addressed; E-mail:  r.patel@griffith.edu.au}\\
\normalsize{or g.pryde@griffith.edu.au}
}

\date{}

\begin{document}


\baselineskip12pt

\twocolumn[
  \begin{@twocolumnfalse}
\maketitle

\begin{sciabstract}
Key to realising quantum computers is minimising the resources required to build logic gates into useful processing circuits. While the salient features of a quantum computer have been shown in proof-of-principle experiments, difficulties in scaling quantum systems have made more complex operations intractable. This is exemplified in the classical Fredkin (controlled-SWAP) gate for which, despite theoretical proposals, no quantum analogue has been realised. By adding control to the SWAP unitary, we use photonic qubit logic to demonstrate the first quantum Fredkin gate, which promises many applications in quantum information and measurement. We implement example algorithms and generate the highest-fidelity three-photon GHZ states to-date. The technique we use allows one to add a control operation to a black-box unitary, something impossible in the standard circuit model. Our experiment represents the first use of this technique to control a two-qubit operation and paves the way for larger controlled circuits to be realised efficiently.
\end{sciabstract}
\end{@twocolumnfalse}
]

\section*{\large\textsf{Introduction}}
One of the greatest challenges in modern science is the realisation of quantum computers\cite{Kok2007,OBrien2009,Ladd2010} which, as they increase in scale, will allow enhanced performance of tasks in secure networking, simulations, distributed computing and other key tasks where exponential speedups are available. Processing circuits to realise these applications are built up from logic gates that harness quantum effects such as superposition and entanglement. At present, even small-scale and medium-scale quantum computer circuits are hard to realise because of the requirement to control enough quantum systems sufficiently well in order to chain together many gates into circuits. One example of this is the quantum Fredkin gate, which requires at least five two-qubit gates\cite{Smolin1996} to be implemented in the standard circuit model. Thus, despite featuring prominently in schemes for quantum computation\cite{Vandersypen2001,Lopez2012,Lanyon2007}, error-correction\cite{Chuang1996,Barenco1997}, cryptography\cite{Buhrman2001,Horn2005,Gottesman2001}, and measurement\cite{Ekert2002,FuiasekFilip2002}, no such gate has been realised to date.

The quantum Fredkin gate, shown in Fig. 1A, is a three-qubit gate whereby, conditioned on the state of the control qubit, the quantum states of the two target qubits are swapped. The original, classical version of the gate first proposed by Fredkin \cite{Fredkin1982} also serves as one of the first examples of a reversible logic operation where the number of bits are conserved and no energy is dissipated as a result of erasure. In the framework of universal quantum computation,  gates are also reversible, so it may seem natural to ask whether it is possible to construct a quantum version of the Fredkin gate. The first design of the quantum Fredkin gate was proposed by Milburn \cite{Milburn1989} and was to use single photons as qubits and cross-Kerr nonlinearities to produce the necessary coherent interactions. Further schemes utilising linear optics developed these ideas further \cite{Chau1995,Smolin1996,Fiurasek2006,Fiurasek2008,Gong2008} by using ancilla photons, interference, and multiple two-qubit\cite{OBrien2003,Pooley2012} and single-qubit gates. However concatenating multiple probabilistic gates in this fashion typically leads to a multiplicative reduction in the overall probability of success of $<1/100$. Hence it would be desirable to be able to construct a quantum Fredkin gate directly without decomposition and avoid the associated resource overhead.

We begin by describing the concept of our experiment. We perform the controlled-SWAP operation by adding control to the SWAP unitary $U_{SWAP}$ applying the technique in Zhou \textit{et al.} \cite{Zhou2011}, to greatly reduce the complexity of quantum circuits. The notion of adding control to a black-box unitary is forbidden or difficult in many architectures\cite{Araujo2014,Thompson2013} -- optics lends itself well to this approach because the optical implementation of the unitary leaves the vacuum state unchanged. Here we utilise this method to simplify a controlled multi-qubit operation.
   \begin{figure*}
   \begin{center}
      \includegraphics[width=\textwidth]{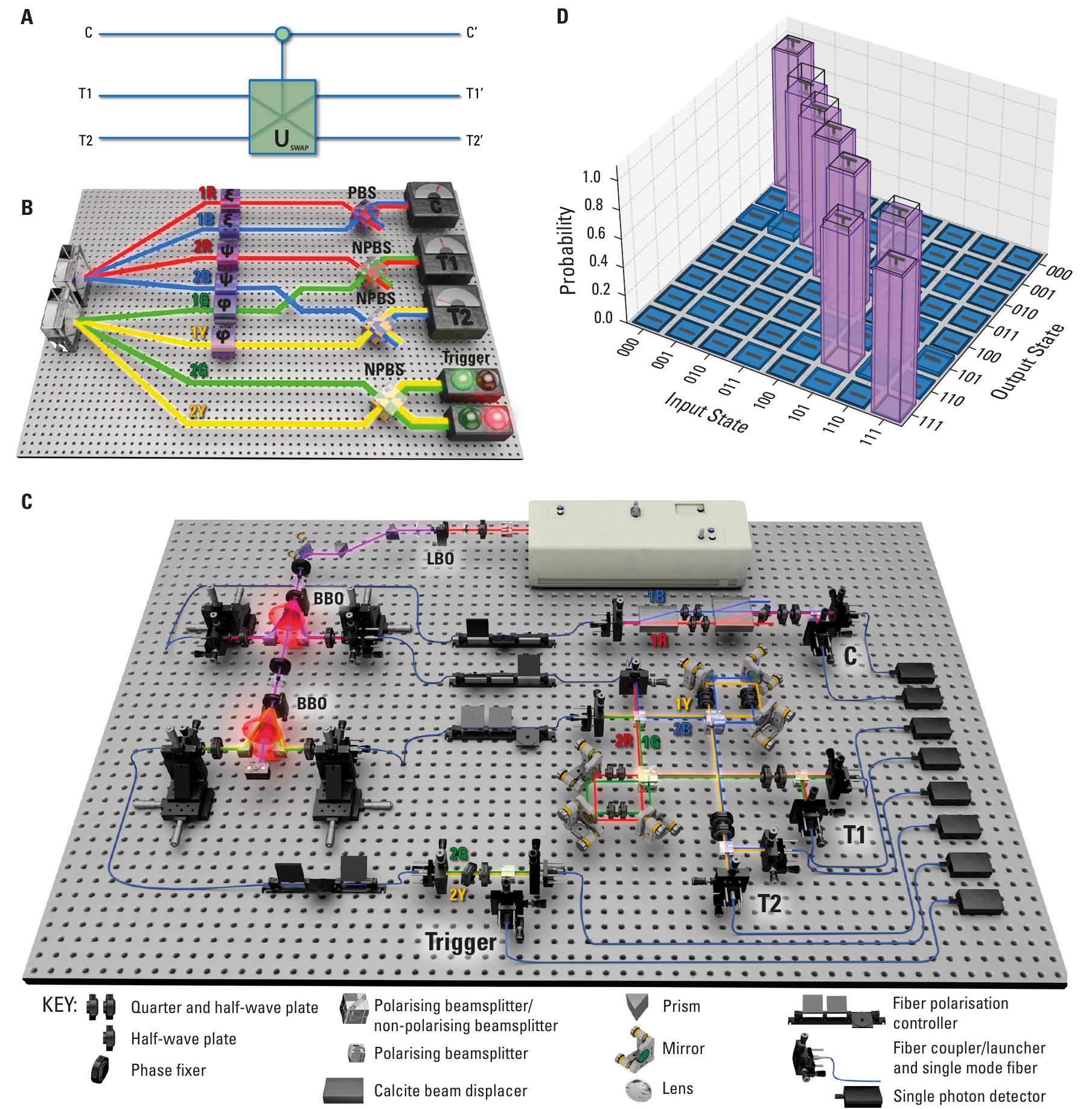}
   \end{center}
\noindent\small{{\bf Fig. 1.} Experimental arrangement and truth table measurements. \textbf{A}, The quantum Fredkin gate circuit. The states of the target qubits are either swapped or not swapped depending on the state of the control qubit. \textbf{B}, Concept of our experiment. Two SPDC photon sources allow production of path entanglement such that modes $R$ and $Y$ are entangled with modes $B$ and $G$. The SWAP operation is carried out on the path modes, depending on the control photon's state,  such that arrival of the control photon indicates a system state of $\alpha|H\rangle^{C}|\psi\rangle^{T1}|\varphi\rangle^{T2} + \beta|V\rangle^{C}|\varphi\rangle^{T1}|\psi\rangle^{T2}$. \textbf{C}, The experimental arrangement. Entangled photons are produced via SPDC  (see Materials and Methods). Entering the gate via single-mode fiber, the two target photons are sent through a PBS. The path-entangled state in Eq. \ref{4GHZ} is produced after each target photon enters a displaced Sagnac interferometer and the which-path information is erased on an NPBS. QWPs and HWPs encode the polarisation state in Eq. \ref{4GHZU}. The control consists of a polarisation beam displacer interferometer. The desired control state is encoded onto modes $1R$ and $1B$ and coherently recombined. A tilted HWP is used to set the phase of the output state. Successful operation is heralded by four-fold coincidence events between the control, target, and trigger detectors. \textbf{D}, Ideal (transparent bars) and measured (solid bars) truth table data for our gate. A total of 620 four-fold events were measured for each of the eight measurements, giving $\left\langle\mathcal{O}\right\rangle = 96\pm4\%$.}
   \end{figure*}
A key idea in our demonstration is to use entanglement in a non-qubit degree of freedom (we use the photon's path mode) to drive the operation of the gate. This path entanglement can be produced in different ways. In our demonstration (Fig. 1B), it is generated from spontaneous parametric down-conversion (SPDC). Given the physical arrangement of the circuit and that we only accept detection events where a single photon is counted at each of the four outputs simultaneously, the optical quantum state produced by SPDC is converted to the required four-photon path-mode entangled state (see Materials and Methods) and has the form
\begin{equation}\label{4GHZ}
 \left(|11\rangle_{B}|11\rangle_{G}|00\rangle_{R}|00\rangle_{Y} + |00\rangle_{B}|00\rangle_{G}|11\rangle_{R}|11\rangle_{Y}\right)/\sqrt{2}
\end{equation}
 where $B$, $R$, $Y$, and $G$ refer to path-modes and, for example, $|11\rangle_{B}$ indicates a photon occupying mode $1B$ and another occupying $2B$. The path-modes are distributed throughout the circuit such that $U_{SWAP}$ is applied only to the $B$ and $G$ modes.  The qubit state is encoded on the polarisation of the photon.  Because the photons are in a spatial superposition, polarisation preparation optics must be applied to both path-modes of each photon. Hence, an arbitrary, separable, three-qubit state $|\xi\rangle|\psi\rangle|\varphi\rangle$ can be prepared as an input to the gate. In particular, the control qubit is encoded on modes $1R$ and $1B$, target 1 is encoded on modes $2R$ and $2B$, and target 2 is encoded on modes $1G$ and $1Y$,  yielding
\begin{equation}\label{4GHZU}
  \left(|\xi\rangle^{C}_{1B}|\psi\rangle^{T1}_{2B}|\varphi\rangle^{T2}_{1G}|H\rangle^{Tr}_{2G} + |\xi\rangle^{C}_{1R}|\psi\rangle^{T1}_{2R}|\varphi\rangle^{T2}_{1Y}|V\rangle^{Tr}_{2Y}\right)/\sqrt{2}
\end{equation}
The two control modes $1R$ and $1B$ are mixed on a polarising beam splitter (PBS), wheras a 50:50 non-polarising beam splitter (NPBS) is used to erase the path information in the target and trigger arms. The SWAP is implemented via rearrangement of the path-modes such that the target modes $2B$ and $1G$ are swapped wheras $2R$ and $1Y$ are not. Successful operation of the gate occurs when photons are detected at the control, target 1, and target 2 detectors (simultaneously with a photon detection at either trigger detector).  The polarisation state of the three-qubit system, given the required modes are occupied, is $\alpha|H\rangle^{C}|\psi\rangle^{T1}|\varphi\rangle^{T2} + \beta|V\rangle^{C}|\varphi\rangle^{T1}|\psi\rangle^{T2}$ as expected from application of the Fredkin gate on the state  $|\xi\rangle^{C}|\psi\rangle^{T1}|\varphi\rangle^{T2}$ where $|\xi\rangle = \alpha|H\rangle + \beta|V\rangle$. Taking into consideration the probability of recording a four-fold coincidence, successful execution of the gate occurs one-sixteenth of
 the time, on average. This can be increased to one-fourth of the time by collecting the target photons from both NPBS outputs.

 \section*{\large\textsf{Results}}
The experimental arrangement of the quantum Fredkin gate is shown in Fig. 1C and consists of three interferometers designed to be inherently phase-stable. Pairs of polarisation entangled photons, produced by two SPDC crystals (see Materials and Methods), impinge on a PBS. Two orthogonally polarised photons, one from each source, are sent to separate displaced Sagnac interferometers. Initially, they are incident on a beam splitter where one half of the interface acts as a PBS and the other half acts as an NPBS. Entering at the PBS side, photons may travel along counterpropagating path modes where the polarisation state $|\psi\rangle$ is encoded onto one mode and the state $|\varphi\rangle$ is encoded on the other. The two paths are then recombined on the NPBS side of the beam splitter where the path information is erased (see Methods and Materials), giving the path-mode entangled state in equation \ref{4GHZ} whilst the polarisation encoding procedure leads to the state in Eq. \ref{4GHZU}. The control of the gate is realised in a polarisation interferometer consisting of two calcite beam displacers. The desired polarisation state of the control is encoded onto modes $1R$ and $1B$, which are coherently recombined in the second beam displacer. Given successful operation (arrival of a photon at the control detector), the preparation of the control photon in $|H\rangle = |1\rangle$ projects the target photons onto path modes $1G$ and $2B$ which undergo SWAP; conversely, preparing $|V\rangle = |0\rangle$ projects the target photons onto path modes $2R$ and $1Y$, which undergo the identity operation. In practice, the trigger arm consists of a half-wave plate (HWP) whose optic axis (OA) is set to $22.5^{\circ}$, producing diagonal $|D\rangle=\frac{1}{\sqrt{2}}(|H\rangle + |V\rangle)$ or anti-diagonal $|A\rangle=\frac{1}{\sqrt{2}}(|H\rangle - |V\rangle)$ polarised photons, and a PBS. Successful operation is heralded by measuring four-fold coincidences across the trigger, control and two target detectors.

The logical operation of the gate was measured by performing eight measurements, one for each of the possible logical inputs. For each input we measure a total of 620 four-fold events distributed across the eight possible output states. Under ideal operation, for a given input, there is a single output. The solid bars in Fig. 1D depict the experimentally measured truth table data, $M_{exp}$ , whereas the transparent bars represent the ideal truth table  $M_{ideal}$. To quantify the mean overlap between $M_{exp}$ and $M_{ideal}$, we calculate $\left\langle\mathcal{O}\right\rangle = Tr \left(M_{exp}M_{ideal}^{T}/M_{ideal}M_{ideal}^{T}\right)= 96\pm4\%$ which confirms excellent performance in the logical basis.  The slight reduction in fidelity is most likely due to the imperfect extinction of our polarisation optics.
  \begin{figure*}
   \begin{center}
      \includegraphics{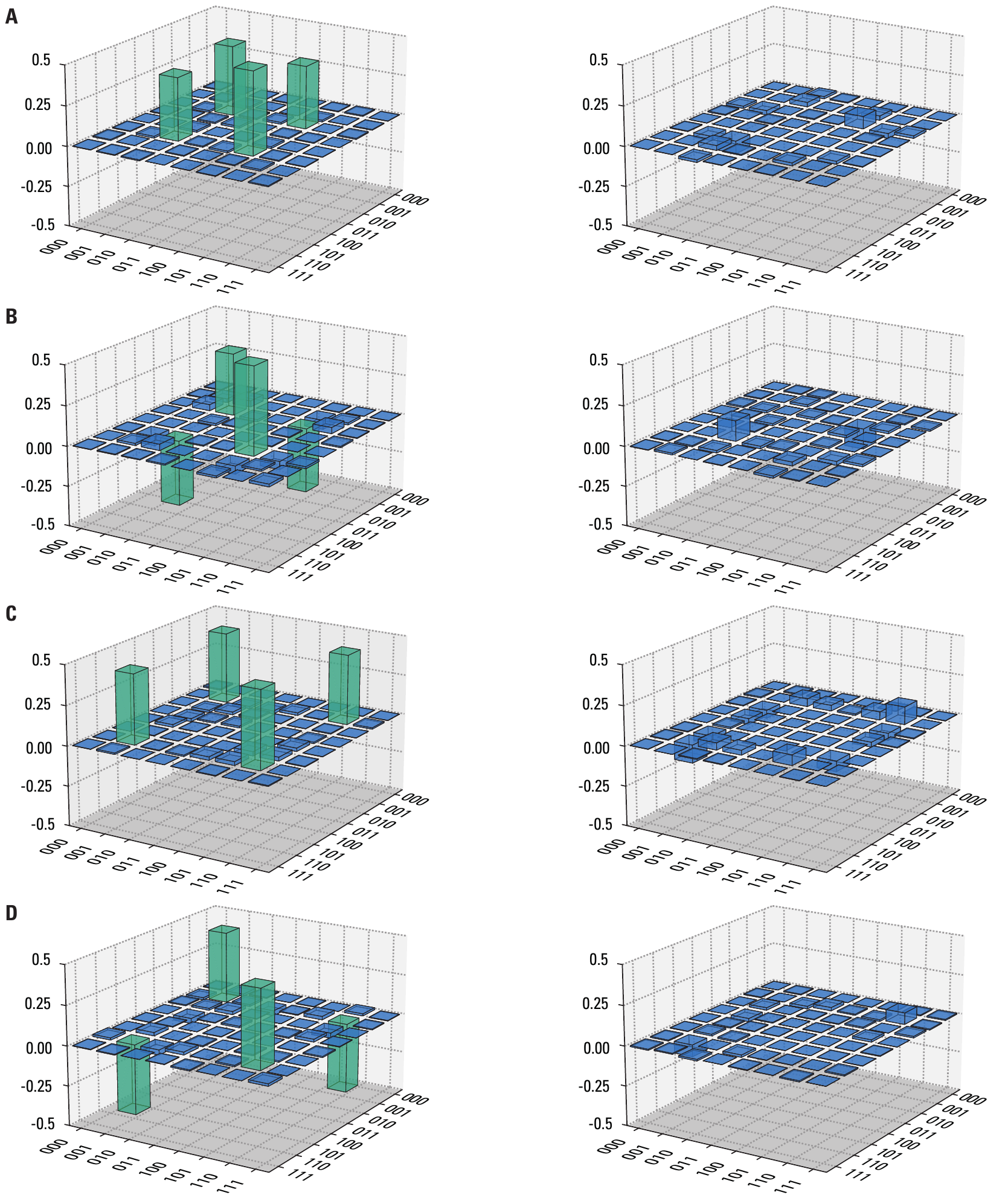}
   \end{center}
   \noindent\small{{\bf Fig. 2.} Real (left) and imaginary (right) parts of the reconstructed density matrices for our four GHZ states. Fidelity and purity were calculated for each state. \textbf{A}, $|GHZ_{1}^{+}\rangle$:  $F = 0.88.\pm 0.01$ and $P = 0.79 \pm 0.02$. \textbf{B},  $|GHZ_{1}^{-}\rangle$: $F = 0.90 \pm 0.01$ and $P = 0.83 \pm 0.02$. \textbf{C},  $|GHZ_{2}^{+}\rangle$: $F = 0.93 \pm 0.01$ and $P = 0.87 \pm 0.02$. \textbf{D}, $|GHZ_{2}^{-}\rangle$: $F = 0.92 \pm 0.01$ and $P = 0.85 \pm 0.02$.}
   \end{figure*}

We demonstrate the full quantum nature of our gate by preparing the control in a superposition $|\xi\rangle = \frac{1}{\sqrt{2}}(|0\rangle \pm |1\rangle)$ which places the gate in a superposition of the SWAP and identity operations. Using our gate, we produce four of the eight maximally entangled three-photon Greenberger-Horne-Zeilinger (GHZ) states, namely
\begin{align}\label{3GHZa}
 \frac{1}{\sqrt{2}}(|0\rangle &\pm |1\rangle)^{C}|1\rangle^{T1}|0\rangle^{T2}\rightarrow |GHZ_1^{\pm}\rangle \nonumber\\
 &= \frac{1}{\sqrt{2}}\left(|0\rangle^{C}|1\rangle^{T1}|0\rangle^{T2} \pm e^{i(\phi + \theta(\vartheta))}|1\rangle^{C}|0\rangle^{T1}|1\rangle^{T2}\right)
\end{align}
and,
 \begin{align}\label{3GHZb}
 \frac{1}{\sqrt{2}}(|0\rangle &\pm |1\rangle)^{C}|0\rangle^{T1}|1\rangle^{T2}\rightarrow |GHZ_2^{\pm}\rangle \nonumber\\
 &= \frac{1}{\sqrt{2}}\left(|0\rangle^{C}|0\rangle^{T1}|1\rangle^{T2} \pm e^{i(\phi + \theta(\vartheta))}|1\rangle^{C}|1\rangle^{T1}|0\rangle^{T2}\right)
\end{align}
Here $\phi$ is a phase shift intrinsic to the gate, and $ \theta(\vartheta)$ is a corrective phase shift that can be applied by tilting a HWP at OA by an angle $\vartheta$, such that $\phi + \theta(\vartheta) = 2n\pi$ (see Materials and Methods). In doing so, we are able to test the coherent interaction of all three qubits in the gate, which is a key requirement for constructing universal quantum computers. For each of the four states in Eqs. \ref{3GHZa} and \ref{3GHZb}, we perform three-qubit quantum state tomography (QST) to fully characterise the state. The control and target qubits are measured independently in the $D/A$ basis, which we denote as $\sigma_x$; in the $R/L$ basis $(\sigma_y)$, where  $|R\rangle=\frac{1}{\sqrt{2}}(|H\rangle + i|V\rangle)$ and  $|L\rangle=\frac{1}{\sqrt{2}}(|H\rangle - i|V\rangle)$; and in the $H/V$ basis $(\sigma_z)$. Therefore full state reconstruction can be carried out by a set of 27 measurements settings $(\sigma_x\sigma_x\sigma_x, \sigma_x\sigma_x\sigma_y...)$ effectively resulting in an over-complete set of 216 projective measurements as each measurement setting has eight possible outcomes. Figure 2 shows the real (left) and imaginary (right) parts of the reconstructed density matrices of the four GHZ states, each of which was calculated from $\sim5000$ four-fold events using a maximum-likelihood algorithm. We measure fidelities and purities of $F = 0.88 \pm 0.01$ and $P = 0.79 \pm 0.02$ for $|GHZ_{1}^{+}\rangle$, $F = 0.90 \pm 0.01$ and $P = 0.83 \pm 0.02$ for $|GHZ_{1}^{-}\rangle$, $F = 0.93 \pm 0.01$, and $P = 0.87 \pm 0.02$ for $|GHZ_{2}^{+}\rangle$, and $F = 0.92\pm 0.01$ and $P = 0.85 \pm 0.02$ for $|GHZ_{2}^{-}\rangle$. The errors were calculated from 500 samples of a Monte-Carlo simulation. These values are most likely limited by imperfect mode overlap at the NPBS in each displaced Sagnac interferometer. Nevertheless, to the best of our knowledge, these values are the highest reported for photonic GHZ states surpassing the previous values reported in Hamel \textit{et al.} \cite{Hamel2014}.
  \begin{figure}[t]
   \begin{center}
   \includegraphics[width=\linewidth]{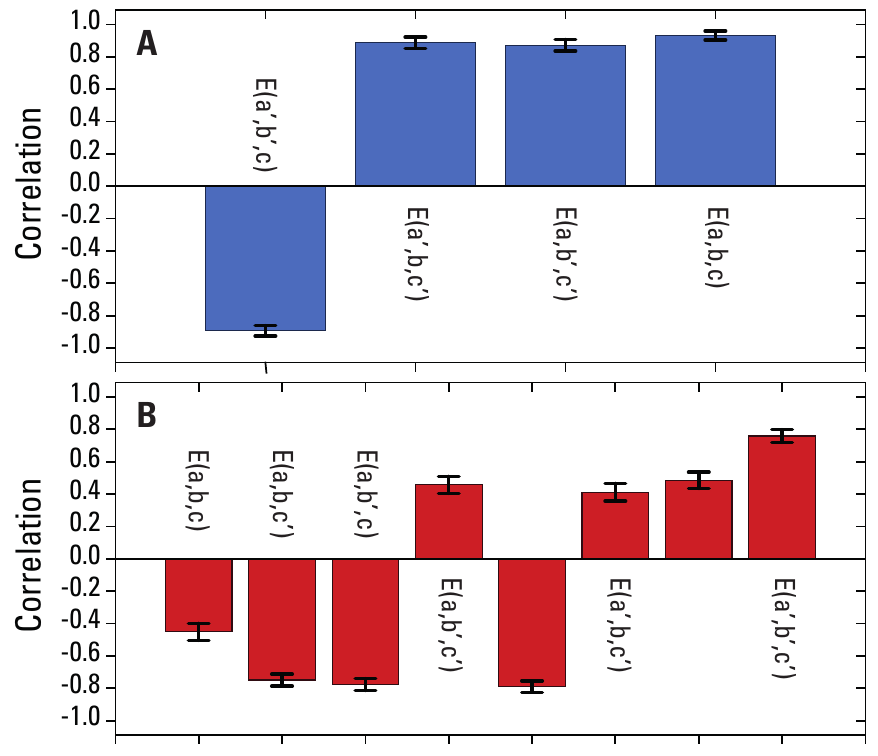}
   \end{center}
   \noindent\small{{\bf Fig. 3.} Measured correlations for violations of Mermin's and Svetlichny's inequalities. \textbf{A},  Mermin's inequality resulting in $S_M = 3.58 \pm 0.06$, a violation by 24 standard deviations. \textbf{B}, Svetlichny's inequality with  $S_{Sv} = 4.88 \pm 0.13$, a violation by 7 standard deviations. Error bars were calculated from Poissonian counting statistics.}
   \end{figure}

We perform further measurements to characterise the quality of the $|GHZ_{2}^{+}\rangle$ state. GHZ states can show a strong contradiction between local hidden-variable theories and quantum mechanics \cite{Pan2000}. Mermin \cite{Mermin1990} derived a Bell-like inequality by imposing locality and realism for three particles, which holds for any local hidden-variable theory
\begin{align}\label{Mermin}
S_M &= |E(a',b,c') + E(a,b',c') + E(a,b,c) - E(a',b',c)|\nonumber\\
 &\leq 2
\end{align}
\begin{figure}[H]
   \begin{center}
      \includegraphics[width=\linewidth]{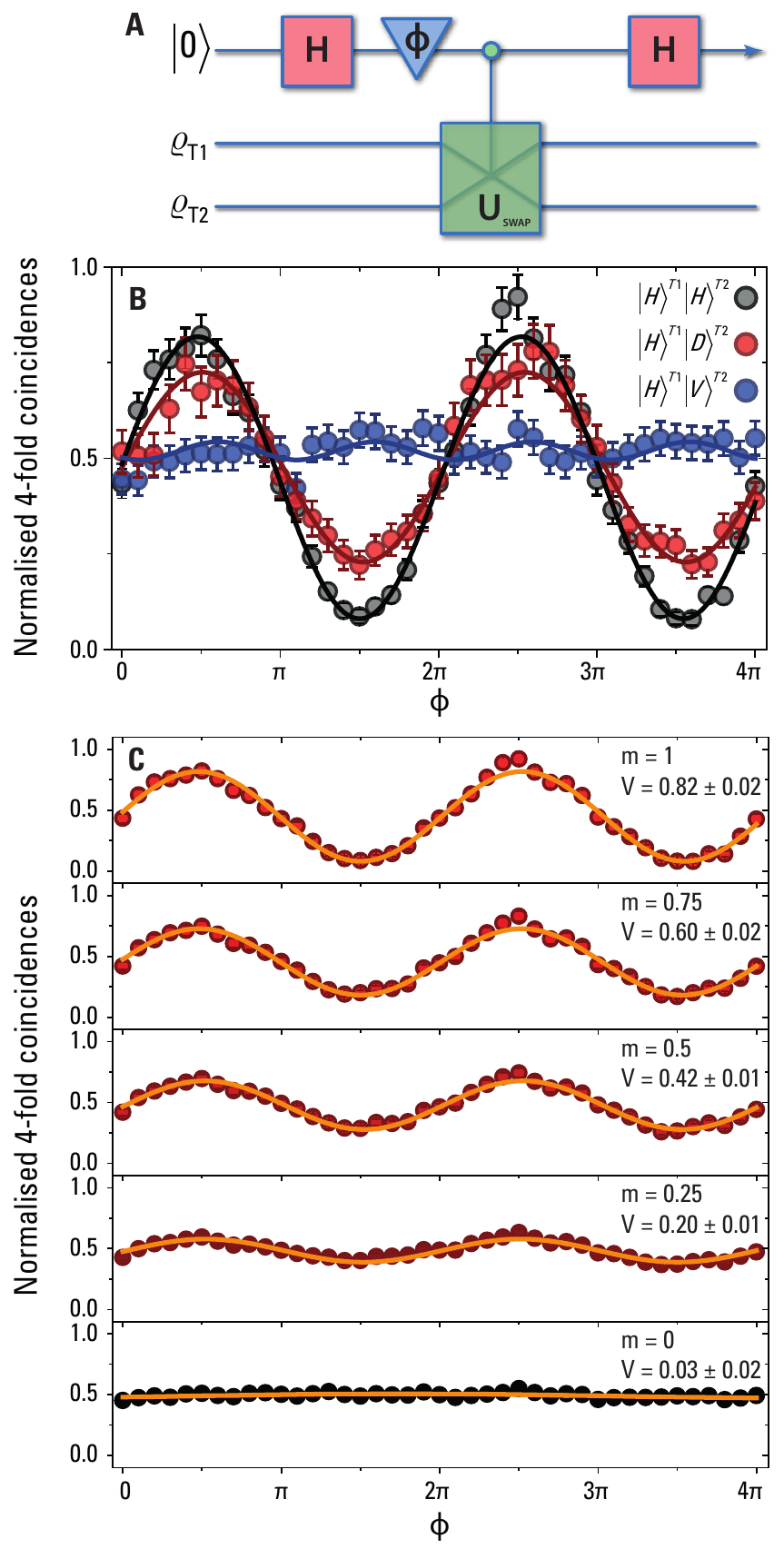}
   \end{center}
   \noindent\small{{\bf Fig. 4.} Estimations of nonlinear functionals of a single-qubit state with the quantum Fredkin gate. \textbf{A},  Circuit diagram of the network. \textbf{B}, Measurements of the overlap of two single qubit states, $|\langle{T1}|{T2}\rangle|^2$.  The fringe visibility or overlap was measured for states $|0\rangle^{T1}|0\rangle^{T2}$ (black), $\frac{1}{\sqrt{2}}\left(|0\rangle \pm 1\rangle\right)^{T1}|0\rangle^{T2}$ (red), and $|0\rangle^{T1}|1\rangle^{T2}$ (blue) with values $0.82 \pm 0.02$, $0.52 \pm 0.02$, and $0.05 \pm 0.01$, respectively. \textbf{C}, Measurements of the state purity.  We measure a visibilities ranging from  $0.82 \pm 0.02$ for a pure state to $0.03 \pm 0.02$ for a maximally mixed state.}
   \end{figure}
This inequality can be violated by performing measurements with settings $a = b = c = \sigma_x$ and $a' = b' = c' = \sigma_y$ with a maximum violation of $S_M = 4$. From the QST of $|GHZ_{2}^{+}\rangle$, 747 of the total 5029 four-fold events can be used to calculate the correlation functions $E$ in Eq. \ref{Mermin}; these results are shown in Fig. 3A. This leads to $S_M = 3.58 \pm 0.06$ which is a violation by 24 standard deviations. The implication of using these particular measurement settings is that the state exhibits genuine tripartite entanglement.

An additional test, namely, the violation of Svetlichny's inequality, is required to test whether the state is capable of displaying tripartite non-locality \cite{Svetlichny1987,Lavoie2009}. Non-local hidden variable theories cannot be ruled out with Mermin's inequality, as they can be violated for arbitrarily strong correlations between two of the three particles. Svetlichny's inequality takes the form
\begin{align}\label{Svet}
S_{Sv} =& |E(a,b,c) + E(a,b,c') + E(a,b',c) - E(a,b',c')\nonumber\\
+& E(a',b,c) - E(a',b,c') - E(a',b',c) - E(a',b',c')|\nonumber\\
&\leq 4
\end{align}
with settings $a = Sv1_\pm$ (where $|Sv1_\pm\rangle = \frac{1}{\sqrt{2}}(|H\rangle \pm e^\frac{i3\pi}{4}|V\rangle)$), $a' = Sv2_\pm$ (where $|Sv2_\pm\rangle = \frac{1}{\sqrt{2}}(|H\rangle \pm e^\frac{i\pi}{4}|V\rangle)$), $b' = c =\sigma_x$, and $b = c' = \sigma_y$. The maximum violation allowed by quantum mechanics is $S_{Sv} = 4\sqrt{2}$. Figure 3B shows the correlations calculated from 2348 four-fold events leading to $S_{Sv} = 4.88 \pm 0.13$, which is a violation by 7 standard deviations.

An application of the quantum Fredkin gate is the direct estimation of non-linear functionals\cite{Ekert2002} of a quantum state, described by a density matrix $\rho$, without recourse to QST. Here $\rho = \varrho_{T1} \otimes \varrho_{T2}$ is the density matrix of two separable subsystems. The circuit we employ is shown in Fig. 4A, where an interferometer is formed using two Hadamard gates and a variable phase shift $\theta(\vartheta)$. This interferometer is coupled to the controlled-SWAP operation of our quantum Fredkin gate such that measuring the control in the logical basis leads to an interference pattern given by $\textrm{Tr}[U_{SWAP}\varrho_{T1} \otimes \varrho_{T2}] = \textrm{Tr}[\varrho_{T1}\varrho_{T2}] = ve^{i\theta(\vartheta)}$. If $\varrho_{T1} \neq \varrho_{T2}$ then measurement of the fringe visibility provides, for pure states, a direct measure of the state overlap $|\langle{T1}|{T2}\rangle|^2$, where $\varrho_{T1} = |{T1}\rangle\langle{T1}|$ and $\varrho_{T2} = |{T2}\rangle\langle{T2}|$. Conversely, if $\varrho_{T1} = \varrho_{T2}$ then the fringe visibility provides an estimate of the length of the Bloch vector ( that is, the purity $P = \textrm{Tr}[\varrho^2]$). We realise the Hadamard operations in Fig. 4A by setting the quarter wave plate (QWP) and HWP combinations to prepare or measure $\sigma_x$.

Figure 4B shows the results of preparing the target qubits in the states $|0\rangle^{T1}|0\rangle^{T2}$, $\frac{1}{\sqrt{2}}\left(|0\rangle + 1\rangle\right)^{T1}|0\rangle^{T2}$, and $|0\rangle^{T1}|1\rangle^{T2}$, corresponding to ideal (measured) overlaps and visibilities of 1 ($0.82 \pm 0.02$), 0.5 ($0.52 \pm 0.02$), and 0 ($0.05 \pm 0.01$), respectively. Although the maximum visibility we are able to measure is limited by the performance of the three interferometers in the circuit, our measurements show a clear reduction in visibility as the single qubit states are made orthogonal. Figure 4C shows the result of setting  $\varrho_{T1} = \varrho_{T2}$. As we increase the degree of mixture (see Materials and Methods), we observe a reduction in visibility from $0.82 \pm 0.02$ for a pure state to $0.03 \pm 0.02$ for a maximally mixed state.

\section*{\large\textsf{Discussion}}
In conclusion, we have used linear optics to perform the first demonstration of the quantum Fredkin gate.  This is achieved by exploiting path-mode entanglement to add control to the SWAP operation. Our implementation has an improved success rate of more than one order of magnitude compared to previous proposals and does not require ancilla photons or decomposition into two-qubit gates. Our gate performs with high accuracy in the logical basis and operates coherently on superposition states. We have used the gate to generate genuine tripartite entanglement with the highest fidelities to date for photonic GHZ states and have implemented a small-scale algorithm to characterise quantum states without QST.

An alternative method for generating the polarisation-path entanglement that drives the gate is the use of C-path gates\cite{Zhou2011} at the input. Our implementation varies from a fully heralded quantum Fredkin gate (see Materials and Methods), which does not require preexisting entanglement; however it demonstrates the key properties of a quantum Fredkin gate. For completely general quantum circuits that incorporate Fredkin (or similar controlled-arbitrary-unitary) gates at arbitrary circuit locations, the C-path methodology may be necessary at the cost of some additional resources and success probability (see Materials and Methods), though we  conjecture that specific circuits comprising multiple Fredkin gates might be optimised using similar techniques to those that allow us to simplify the Fredkin down from a circuit of five two-qubit gates.  Nevertheless, for small algorithms or operations and whenever possible, it is significantly favourable to directly generate path entanglement.

The quantum Fredkin gate has many applications across quantum information processing. Our demonstration should stimulate the design and implementation of even more complex quantum logic circuits. Later we became aware of related work carried out by Takeuchi\cite{Takeuchi2015}.

\section*{\large\textsf{Materials and Methods}}

\paragraph*{Source\\}
Our source consisted of a $150\textrm{ fs}$ pulsed Ti-Sapphire laser operating at a rate of  $80\textrm{ MHz}$ and at a wavelength of $780\textrm{ nm}$, which was frequency doubled using a $2\textrm{ mm}$ LBO crystal. Two dispersion-compensating ultrafast prisms spatially filter any residual  $780\textrm{ nm}$ laser light. The frequency- doubled light (with power $100\textrm{ mW}$) pumped two $2\textrm{ mm}$ type-II $\beta$ barium borate (BBO) crystals in succession. Entangled photons, generated via SPDC, were collected at the intersection of each set of emission cones. They then encountered an HWP with its OA at $45^{\circ}$ and an additional $1\textrm{ mm}$ type-II BBO crystal used to compensate for spatial and temporal walk-offs. The single photons were coupled into single-mode fiber and delivered to the gate. This configuration gave, on average, a four-fold coincidence rate of 2.2 per minute at the output of the gate.
\paragraph*{Entangled state preparation\\}
Each SPDC source emitted pairs of entangled photons of the form $|\psi^{+}_{1}\rangle = \frac{1}{\sqrt{2}}\left(|H\rangle_{1B}|V\rangle_{2B}+|V\rangle_{1R}|H\rangle_{2R}\right)$ and $|\psi^{+}_{2}\rangle = \frac{1}{\sqrt{2}}\left(|H\rangle_{1Y}|V\rangle_{2Y}+|V\rangle_{1G}|H\rangle_{2G}\right)$. Polarisation optics were used to distribute the path-modes throughout the circuit and thus convert this state into the path-entangled states $|\psi^{+}_{1}\rangle = \frac{1}{\sqrt{2}}\left(|1\rangle_{1B}|1\rangle_{2B}|0\rangle_{1R}|0\rangle_{2R}+|0\rangle_{1B}|0\rangle_{2B}|1\rangle_{1R}|1\rangle_{2R}\right)$ and \\$|\psi^{+}_{2}\rangle = \frac{1}{\sqrt{2}}\left(|1\rangle_{1Y}|1\rangle_{2Y}|0\rangle_{1G}|0\rangle_{2G}+|0\rangle_{1Y}|0\rangle_{2Y}|1\rangle_{1G}|1\rangle_{2G}\right)$. Path modes from $|\psi^{+}_{1}\rangle$ and $|\psi^{+}_{2}\rangle$ were combined on a PBS (Fig 1C, PBS with outputs $2R$, $1G$, $1Y$, and $2B$); along with post-selection of four-fold coincidence events at the outputs of the control, target, and trigger outputs, this led to Eq. \eqref{4GHZ} in the main text. Each qubit was encoded using photon polarisation: using Eq. \eqref{4GHZ}, considering that each photon exists in a superposition of path-modes and omitting the unoccupied modes, an arbitrary polarisation state can be encoded onto each qubit by performing a local unitary operation on each mode, giving equation \eqref{4GHZU}. The state encoding was performed inside the beam displacer (control qubit) and displaced Sagnac (target qubits) interferometers.
\paragraph*{Tuning the phase\\}
The phase was tuned by tilting an HWP set to its OA. To set the correct phase for each of the four GHZ states, we varied the tilt of the HWP and measured fringes in the four-fold coincidences with our measurement apparatus in the $\sigma_x\sigma_y\sigma_y$ basis. For the $|GHZ_{1,2}^{+}\rangle$ $\left(|GHZ_{1,2}^{-}\rangle\right)$ we set the tilt to maximise (minimise) the occurrence of the $|DRR\rangle$, $|DLL\rangle$, $|ARL\rangle$, and $|ALR\rangle$ events.
\paragraph*{Mixed state preparation\\}
The mixed states of the form $\varrho = m|0\rangle\langle{0}| + \frac{(1-m)}{2}\left(|0\rangle\langle{0}| + |1\rangle\langle{1}|\right)$ were obtained by measuring output statistics for a combination of pure input states. The input states of the target were prepared, in varying proportions given by the parameter $m$, as  $0.25 (1 + m)^2|0\rangle^{T1}|0\rangle^{T2}$, $0.25(1 - m^2)|0\rangle^{T1}|1\rangle^{T2}$, $0.25(1 - m^2)|1\rangle^{T1}|0\rangle^{T2}$, and $0.25 (1 - m)^2|1\rangle^{T1}|1\rangle^{T2}$.  The aggregated data resulted in a fringe pattern which reflects the purity of the mixed single-qubit state.
\paragraph*{Erasing the which-path information\\}
Generation of path-mode entanglement and successful operation of the gate in the quantum regime relied on the erasure of the which-path information in the two displaced Sagnac interferometers. We tested this by performing a Hong-Ou-Mandel (HOM) two-photon interference measurement after each interferometer. After overlapping path modes $2R$ and $1G$ on an NPBS, an HWP with its OA set to $22.5^{\circ}$ rotated the polarisation of the photons to $|D\rangle$ and $|A\rangle$, respectively. Sending these photons into the same port of a PBS led to bunching at the output if the path-modes were indistinguishable. Doing the same for modes $2B$ and $1Y$ gave two separate HOM dips (see Materials and Methods) with visibilities of $90 \pm 5\%$ and $91 \pm 6\%$.
\paragraph*{Heralding the quantum Fredkin gate\\}
In order to use a quantum Fredkin gate as part of a much larger quantum circuit (with gates in series), it is preferable for the gate to be heralded. Realising our gate in this manner involves adding C-path gates\cite{Zhou2011} to each input. For the best probability of success $P_{success}$, each C-path gate requires two heralded C-NOT gates\cite{Pittman2001} which, in turn, require two entangled pair ancillae. Execution of the C-path gate succeeds with $P_{success} = (1/4)^2$\cite{Zhou2011,Pittman2001}. C-path gates are not a necessity at the output if successful execution is heralded by non-detections at the relevant NPBS ports, at an additional probability cost of factor $1/4$.

\bibliography{fredkin}
\bibliographystyle{ScienceAdvances}

\noindent \textbf{Acknowledgements:}

We thank Sergei Slussarenko, Allen Boston, and Kavan Modi for useful discussions.\\
\noindent \textbf{Funding:} This work was supported by the Australian Research Council Centre of Excellence for Quantum Computation and Communication Technology (Project number CE110001027).\\
\noindent \textbf{Author Contributions} T.C.R., F.F., and G.J.P. conceptualised the scheme. R.B.P., J.H., and F.F. designed and constructed the experiment. R.B.P. and J.H. performed the experiment and data analysis with input from G.J.P. R.B.P. wrote the manuscript with input from all authors. G.J.P. oversaw the work.\\
\noindent \textbf{Competing Interests} The authors declare that they have no competing financial interests.\\
\noindent \textbf{Data and materials availability:} All data needed to evaluate the conclusions in the paper are present in the paper itself and in the Supplementary Materials or are available upon request from the authors.

\noindent \textbf{Supplementary Materials:}\\
Section S1. Erasing the which-path information\\
Section S2. Generation of three-photon GHZ states\\
Fig. S1. HOM dip measurements

\clearpage
  

\section*{Supplementary material}

## S1. Erasing the which-path information

Generation of path-mode entanglement, and successful operation of the gate in the quantum regime, relies on the erasure of the which-path information in the two displaced Sagnac interferometers. We test this by performing a Hong-Ou-Mandel (HOM) two-photon interference measurement after each interferometer. After overlapping path modes $2R$ and $1G$ on an NPBS a HWP with its OA set to 22.5° rotates the polarisation of the photons to $|D\rangle$ and $|A\rangle$, respectively. Sending these photons into the same port of a PBS leads to bunching at the output if the path-modes are indistinguishable. Doing the same for modes $2B$ and $1Y$ gives two separate HOM dips with visibilities 90±5% and 91±6%. In the main article, the HOM dips were measured simultaneously for each displaced Sagnac interferometer where path modes $2R$ and $1G$, and $2B$ and $1Y$ are overlapped on a non-polarising beamsplitter (NPBS). The temporal delay in arrival times of the photons was varied by scanning the position of the input coupler for path modes $1G$ and $1Y$. The variation in accumulated four-fold events as a function of temporal delay results in a dip with a visibility defined as $V = (C_{max} - C_{min})/C_{max}$, where $C_{max}$ and $C_{min}$ are the maximum and minimum number of four-fold events, respectively. In Fig S1 we measure $V=90\pm5\%$ for modes $2R$ and $1G$ and $V=91\pm6\%$ for modes $2B$ and $1Y$, confirming a high degree of indistinguishability.

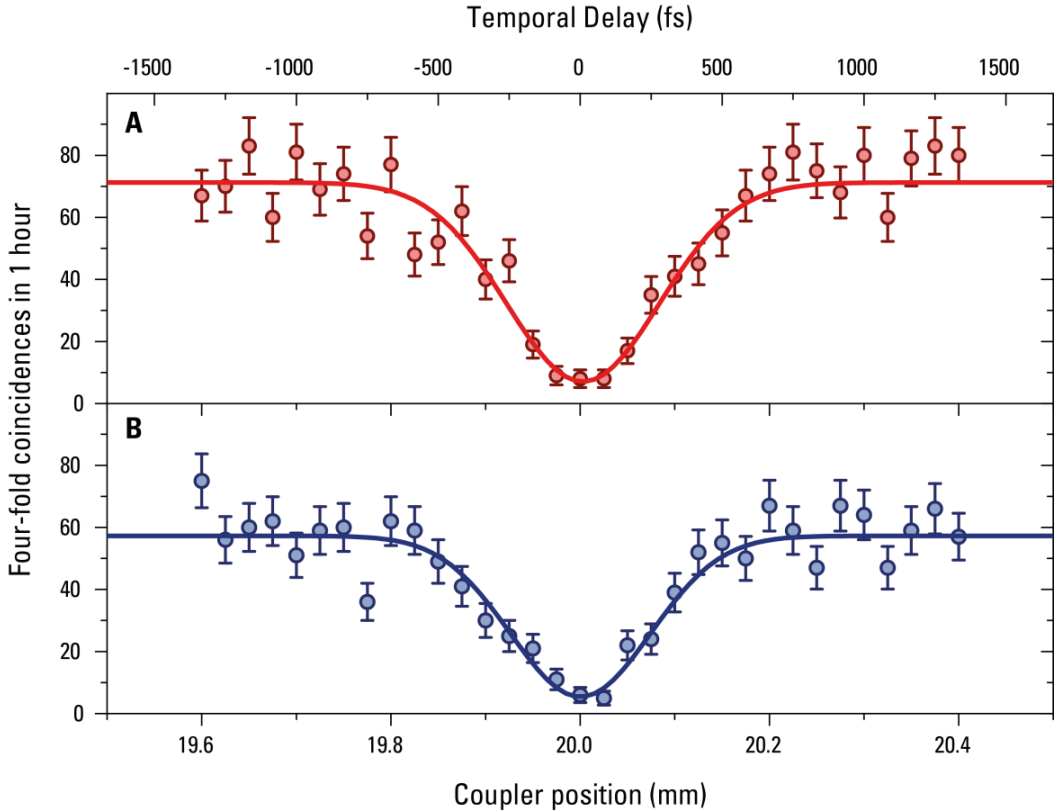

**Fig. S1. HOM dip measurements testing the indistinguishability of path-modes after each displaced Sagnac interferometer.** (**A**), Overlap of $2R$ and $1G$ with V=90±5%, and (**B**), overlap of $2B$ and $1Y$ with V=91±6%.

## S2. Generation of three-photon GHZ states

Here we outline how our quantum Fredkin gate generates four of the eight maximally entangled three-photon GHZ states. The general four-photon state prior to state preparation is

$$\left(|\xi\rangle_{1B}^{C}|\psi\rangle_{2B}^{T1}|\varphi\rangle_{1G}^{T2}|H\rangle_{2G}^{Tr}+|\xi\rangle_{1R}^{C}|\psi\rangle_{2R}^{T1}|\varphi\rangle_{1Y}^{T2}|V\rangle_{2Y}^{Tr}\right)\Big/\sqrt{2} \tag{S1}$$

The photons in modes $2G$ and $2Y$ pass through a HWP with its OA set to 22.5° and then impinge on a PBS leading to

$$\frac{|\xi\rangle_{1B}^{C}|\psi\rangle_{2B}^{T1}|\varphi\rangle_{1G}^{T2}|H\rangle_{2G}^{Tr}}{\sqrt{4}}+\frac{|\xi\rangle_{1B}^{C}|\psi\rangle_{2B}^{T1}|\varphi\rangle_{1G}^{T2}|V\rangle_{2G}^{Tr}}{\sqrt{4}}+$$
$$\frac{|\xi\rangle_{1R}^{C}|\psi\rangle_{2R}^{T1}|\varphi\rangle_{1Y}^{T2}|H\rangle_{2Y}^{Tr}}{\sqrt{4}}-\frac{|\xi\rangle_{1R}^{C}|\psi\rangle_{2R}^{T1}|\varphi\rangle_{1Y}^{T2}|V\rangle_{2Y}^{Tr}}{\sqrt{4}} \tag{S2}$$

In order to prepare the gate in a superposition of the SWAP and identity operations the control qubit is set to $|D\rangle^{C}=\frac{1}{\sqrt{2}}\left(|H\rangle^{C}+|V\rangle^{C}\right)$ giving

$$\frac{|H\rangle_{1B}^{C}|\psi\rangle_{2B}^{T1}|\varphi\rangle_{1G}^{T2}|H\rangle_{2G}^{Tr}}{\sqrt{8}}+\frac{|V\rangle_{1B}^{C}|\psi\rangle_{2B}^{T1}|\varphi\rangle_{1G}^{T2}|H\rangle_{2G}^{Tr}}{\sqrt{8}}$$
$$+\frac{|H\rangle_{1B}^{C}|\psi\rangle_{2B}^{T1}|\varphi\rangle_{1G}^{T2}|V\rangle_{2G}^{Tr}}{\sqrt{8}}+\frac{|V\rangle_{1B}^{C}|\psi\rangle_{2B}^{T1}|\varphi\rangle_{1G}^{T2}|V\rangle_{2G}^{Tr}}{\sqrt{8}}$$
$$+\frac{|H\rangle_{1R}^{C}|\psi\rangle_{2R}^{T1}|\varphi\rangle_{1Y}^{T2}|H\rangle_{2Y}^{Tr}}{\sqrt{8}}+\frac{|V\rangle_{1R}^{C}|\psi\rangle_{2R}^{T1}|\varphi\rangle_{1Y}^{T2}|H\rangle_{2Y}^{Tr}}{\sqrt{8}}$$
$$-\frac{|H\rangle_{1R}^{C}|\psi\rangle_{2R}^{T1}|\varphi\rangle_{1Y}^{T2}|V\rangle_{2Y}^{Tr}}{\sqrt{8}}-\frac{|V\rangle_{1R}^{C}|\psi\rangle_{2R}^{T1}|\varphi\rangle_{1Y}^{T2}|V\rangle_{2Y}^{Tr}}{\sqrt{8}} \tag{S3}$$

In the control arm of our experiment, the interferometer is arranged such that terms with $|H\rangle_{1B}^{C}$ or $|V\rangle_{1R}^{C}$ in equation (S3) are rejected. Rejecting these terms and swapping modes $2B$ and $1G$ gives

$$\frac{|H\rangle_{1R}^{C}|\psi\rangle_{2R}^{T1}|\varphi\rangle_{1Y}^{T2}|H\rangle_{2Y}^{Tr}}{\sqrt{4}}+\frac{|V\rangle_{1B}^{C}|\varphi\rangle_{1G}^{T1}|\psi\rangle_{2B}^{T2}|H\rangle_{2G}^{Tr}}{\sqrt{4}}$$
$$-\frac{|H\rangle_{1R}^{C}|\psi\rangle_{2R}^{T1}|\varphi\rangle_{1Y}^{T2}|V\rangle_{2Y}^{Tr}}{\sqrt{4}}+\frac{|V\rangle_{1B}^{C}|\varphi\rangle_{1G}^{T1}|\psi\rangle_{2B}^{T2}|V\rangle_{2G}^{Tr}}{\sqrt{4}} \tag{S4}$$

Consequently from equation (S4) detection of a trigger photon in state $|H\rangle$ or $|V\rangle$ will result in two states with a relative phase difference of $\pi$.

$$|H\rangle^{Tr}:\frac{|H\rangle_{1R}^{C}|\psi\rangle_{2R}^{T1}|\varphi\rangle_{1Y}^{T2}}{\sqrt{2}}+\frac{|V\rangle_{1B}^{C}|\varphi\rangle_{1G}^{T1}|\psi\rangle_{2B}^{T2}}{\sqrt{2}} \tag{S5}$$

$$|V\rangle^{Tr} : \frac{|H\rangle^C_{1R}|\psi\rangle^{T1}_{2R}|\varphi\rangle^{T2}_{1Y}}{\sqrt{4}} - \frac{|V\rangle^C_{1B}|\varphi\rangle^{T1}_{1G}|\psi\rangle^{T2}_{2B}}{\sqrt{2}} \qquad (S6)$$

As we detect both polarisations of the trigger photon, it is necessary to perform a classical phase rotation to the coincidence data corresponding to the detection of $|V\rangle^{Tr}$. The two sets of coincidence data are then combined. After erasing the which-path information, setting $|\psi\rangle = |V\rangle$ and $|\varphi\rangle = |H\rangle$ and recalling that $|H\rangle = |1\rangle$ and $|V\rangle = |0\rangle$ we obtain the three-photon GHZ state

$$|GHZ_1^+\rangle = \frac{\left(|0\rangle^C|1\rangle^{T1}|0\rangle^{T2} + |1\rangle^C|0\rangle^{T1}|1\rangle^{T2}\right)}{\sqrt{2}}. \qquad (S7)$$

Alternatively, setting $|\psi\rangle = |H\rangle$ and $|\varphi\rangle = |V\rangle$ produces

$$|GHZ_2^+\rangle = \frac{\left(|0\rangle^C|0\rangle^{T1}|1\rangle^{T2} + |1\rangle^C|1\rangle^{T1}|0\rangle^{T2}\right)}{\sqrt{2}}. \qquad (S8)$$

Preparing the control in $|A\rangle^C = \frac{1}{\sqrt{2}}\left(|H\rangle^C - |V\rangle^C\right)$ and setting $|\psi\rangle = |V\rangle$ and $|\varphi\rangle = |H\rangle$ gives

$$|GHZ_1^-\rangle = \frac{\left(|0\rangle^C|1\rangle^{T1}|0\rangle^{T2} - |1\rangle^C|0\rangle^{T1}|1\rangle^{T2}\right)}{\sqrt{2}} \qquad (S9)$$

while for $|\psi\rangle = |H\rangle$ and $|\varphi\rangle = |V\rangle$

$$|GHZ_2^-\rangle = \frac{\left(|0\rangle^C|0\rangle^{T1}|1\rangle^{T2} - |1\rangle^C|1\rangle^{T1}|0\rangle^{T2}\right)}{\sqrt{2}} \qquad (S10)$$


\end{document}